# Production of photoionized plasmas in the laboratory using X-ray line radiation

S White[1], R Irwin[1], R Warwick[1], G Gribakin[1], G Sarri[1], F P Keenan[1], D Riley[1], S J Rose[2,3], E G Hill[2], G J Ferland[4], B Han[5], F Wang[6], and G Zhao[6]

[1]*School of Mathematics and Physics, Queen's University Belfast, University Road, Belfast BT7 1NN, UK*

[2]*Plasma Physics Group, Blackett Laboratory, Prince Consort Road, London SW7 2AZ, UK*

[3]*Clarendon Laboratory, University of Oxford, Parks Road, Oxford OX1 3PU, UK*

[4]*Department of Physics and Astronomy, University of Kentucky, Lexington, KY 40506, USA*

[5]*Department of Astronomy, Beijing Normal University, Beijing 100875, People's Republic of China*

[6]*Key Laboratory of Optical Astronomy, National Astronomical Observatories, Chinese Academy of Sciences, Beijing 100012, People's Republic of China*

**Abstract**
In this paper we report the experimental implementation of a theoretically-proposed technique for creating a photoionized plasma in the laboratory using X-ray line radiation. Using a Sn laser-plasma to irradiate an Ar gas target, the photoionization parameter , $\xi = 4\pi F/N_e$, reached values of order 50 erg cm s$^{-1}$ , where F is the radiation flux in erg cm$^{-2}$s$^{-1}$. The significance of this is that this technique allows us to mimic effective spectral radiation temperatures in excess of 1keV. We show that our plasma starts to be collisionally dominated before the peak of the X-ray drive. However, the technique is extendable to higher energy laser systems to create plasmas with parameters relevant to benchmarking codes used to model astrophysical objects.

## I. Introduction

Photoionized plasmas are thought to be regularly associated with accretion-powered astrophysical objects [1-6], where the radiation field is sufficiently intense that photoexcitation and ionization rates are high relative to electron collisional excitation and ionization rates. The distribution of ionization is characterised by the photoionization parameter $\xi$ which may reach $\simeq$ 1000 erg cm s$^{-1}$ or greater [e.g. 7]. There have been several attempts to create such photoionized plasmas in the laboratory, to allow plasma modelling codes to be benchmarked against well-diagnosed experimental data. However, as the electron density is generally much larger than the astrophysical counterpart, so too must be the X-ray flux to generate high values of $\xi$. This can be difficult, even on major facilities. Foord *et al.* [8] created a photoionized plasma using X-rays from the Z Machine pinch at the Sandia National Laboratories, the X-ray flux acting both to heat and decompress the original foil, and to subsequently bathe the resulting plasma in a radiation field to achieve values of $\xi \simeq 25$ erg cm s$^{-1}$. More recently, Fujioka *et al.* [9] generated a photoionized plasma with the high-power GEKKO-XII laser facility with $\xi \simeq 6$ erg cm s$^{-1}$ and an effective radiation temperature of ~0.5 keV albeit for < 200 ps. Loisel *et al*. [10] recently achieved $\xi \simeq 20$ erg cm s$^{-1}$ on the Z Machine pinch.

We have used an experimental technique based on a proposal by Hill and Rose [11] that replaces the usual broad-band, quasi-black-body source with an intense *line* radiation source. As well as providing a high X-ray flux, this allows keV radiation to dominate over the softer X-rays and thus mimics the effect of a higher spectral temperature on the photoionization of the target gas, where in the limit inner-shell photoionization should dominate over outer-shell photoionization. This is an advantage if we are to use such experiments to benchmark codes under relevant conditions.

## II. Experimental set-up

Our experiment was carried out on the VULCAN laser at the Rutherford Appleton Laboratory. The target, illustrated in Fig. 1, was an Ar-filled gas cell, made from Al, with a window on each face. The 'east' window consisted of a 18.8 µm polypropylene (CH) foil coated on the outside with 218 nm of Sn for full shots This window was fitted to a circular re-entrant plug that allowed the window to be further inside the gas cell. A 500 µm diameter hole in the 500 µm thick base of the plug also served to partially collimate the X-ray flux onto the gas fill. The 'south' face was covered with a 20 $\mu$m Kapton window that allowed the emission of fluorescent X-rays (Ar K-α and K-β at ≃ 3 keV) from the heated gas to be imaged by a spherical Mica crystal with spatial imaging in the direction normal to the focal spot of the heating beams. The 'west' window, opposite the heating beams and viewing into the recessed aperture, was also 20 $\mu$m of Kapton and allowed the L-shell drive radiation to be monitored by a Si(111) spectrometer [12] and X-ray streak camera. A pinhole camera fitted with a 10 µm pinhole, with a magnification of 2.56±0.03, monitored the source size in the keV region. Remaining windows were also Kapton and provided lines-of-sight for target alignment.

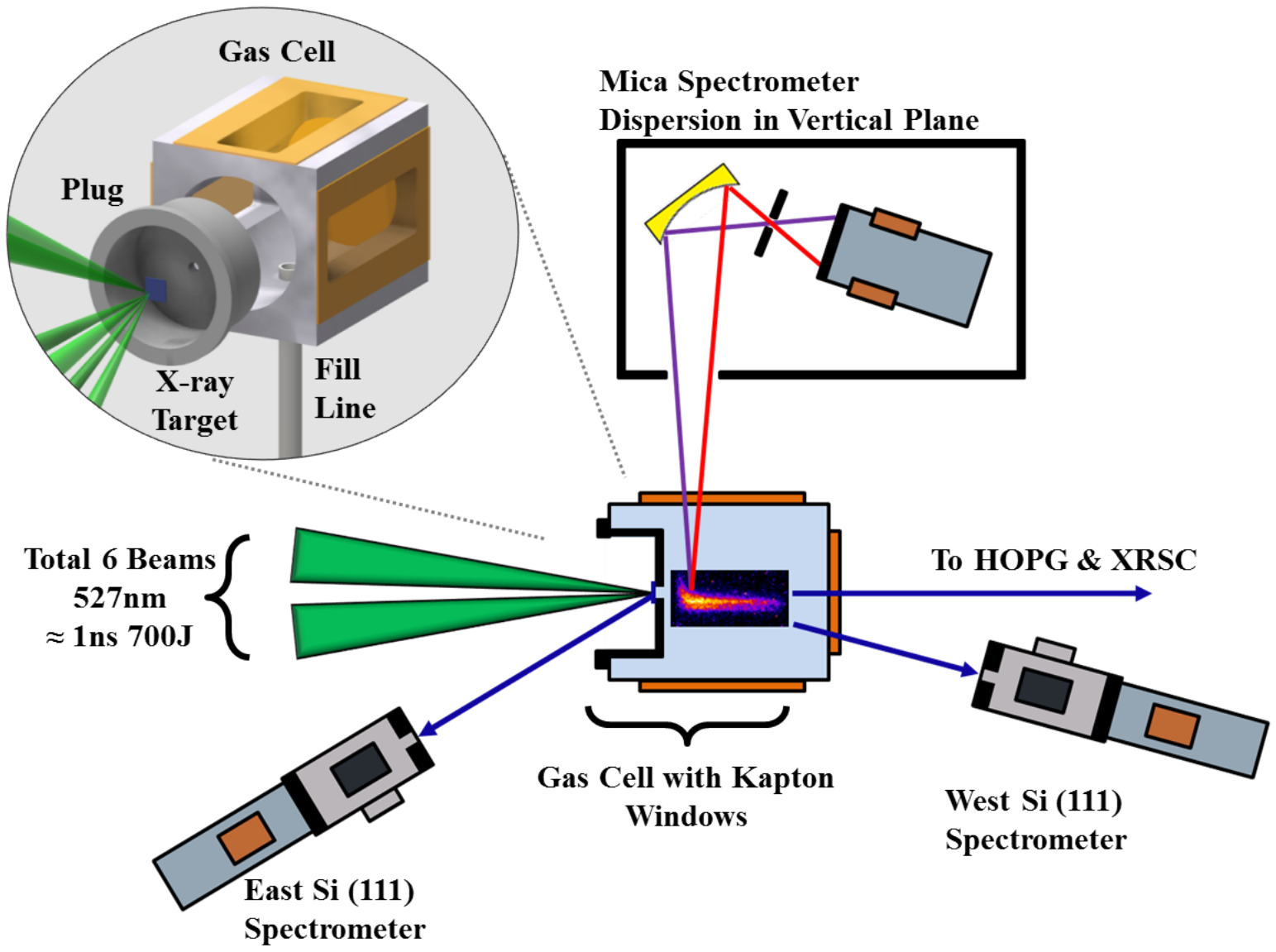

FIG. 1. Inset: Schematic of the gas cell. Main: Experimental schematic showing the drive laser beams incident upon a thin layer of Sn coated on top of CH. The X-ray target was placed over a 0.5 mm opening made in a re-entrant plug which fitted inside the 'East' Window of the gas cell. This allowed observation of the K-shell florescence close to the X-ray source by a spherically curved Mica crystal spectrometer through a Kapton window mounted on the south side of the gas cell. The dispersion plane was vertical, allowing us to spectrally resolve and to spatially resolve X-ray emission along a direction normal to the laser focal plane. Two flat silicon crystal X-ray spectrometers were used to spectrally characterise the X-ray target, with one viewing the laser irradiated side of the target and the other through the target and a second Kapton window. In addition, through this window a streak camera coupled to a flat HOPG crystal measured the temporal profile.

A hypodermic needle inserted into the base of the gas cell acted as a fill line and mounting post. The gas cell was illuminated with 6 overlapping laser pulses (up to 850 J at 527 nm) in a 300–500 μm focal spot with a 1.5 ns flat top temporal profile. This created a source of Sn L-shell line radiation in the 3.4–4.3 keV range which photoionized the Ar gas. In Figure 2 we plot sample spectra from the Si crystal spectrometers, with a wide spectral range explored by moving the spectrometers over a series of shots. Analysis of the spectra transmitted through the CH foil side of the target allowed us to estimate that a total of ≈ 2.0±0.5% of the laser energy converted to L-shell radiation on a typical shot.

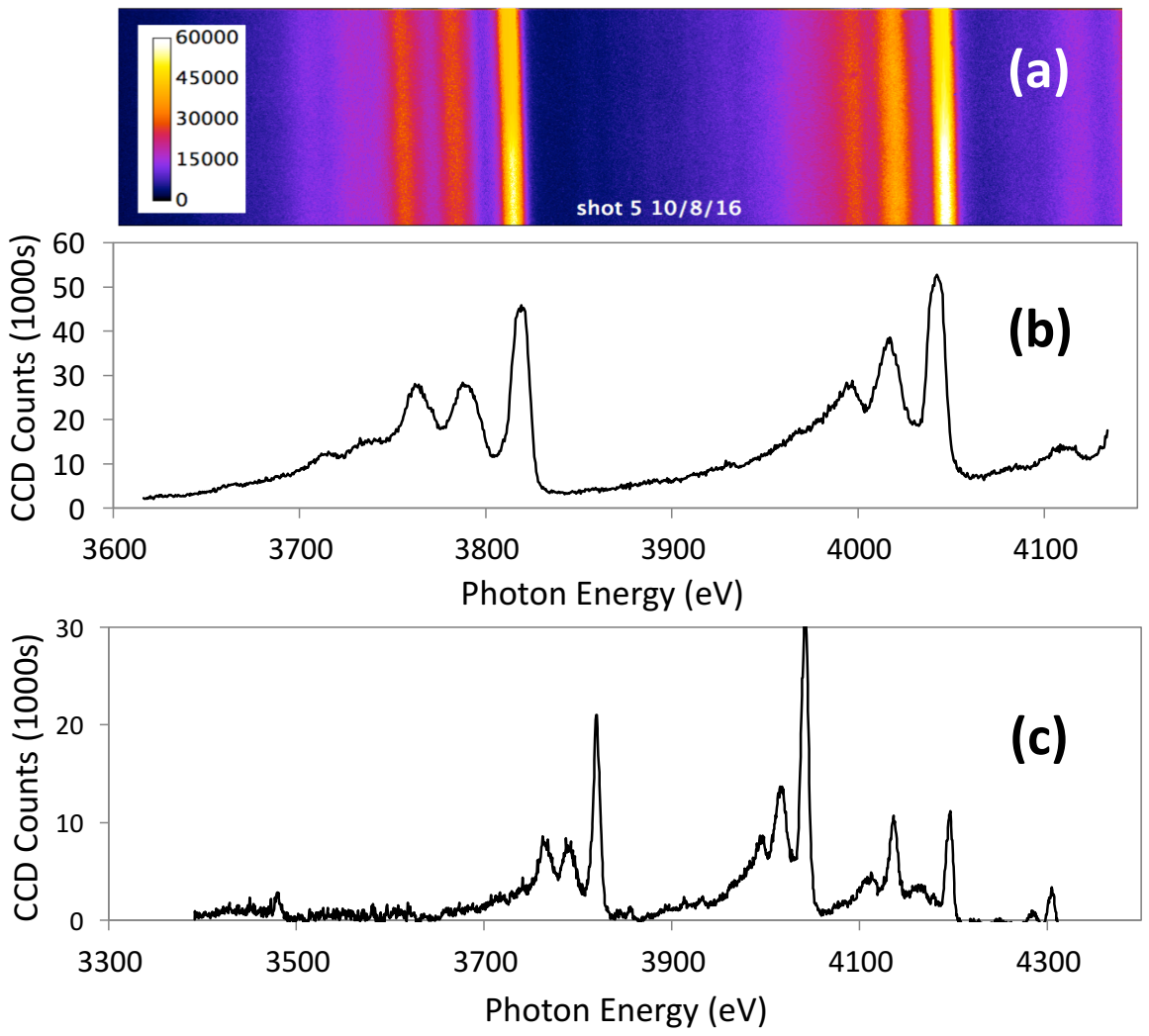

FIG.2. (a) Raw spectrum for a typical shot taken with a flat Si crystal spectrometer. (b) Lineout from the raw image. The energy calibration of the spectrometer was performed using a test spectrum from a Ca target with well-known He-like and Li-like lines. (c) Lineout combining several shots with various Bragg angles, and scaled according to laser energy, to assemble the whole L-band spectrum from Sn.

The contribution of lower energy (sub-keV to keV) radiation from the Sn foil was monitored in a narrow range centred around 1.5 keV using a TAP crystal. Based on these measurements and earlier work with Pd foils [13] we have estimated the flux of radiation that will pass through the CH filtering. This is about 20±5% of the L-shell flux and  corresponds to an effective black-body radiation temperature of 175±10eV.

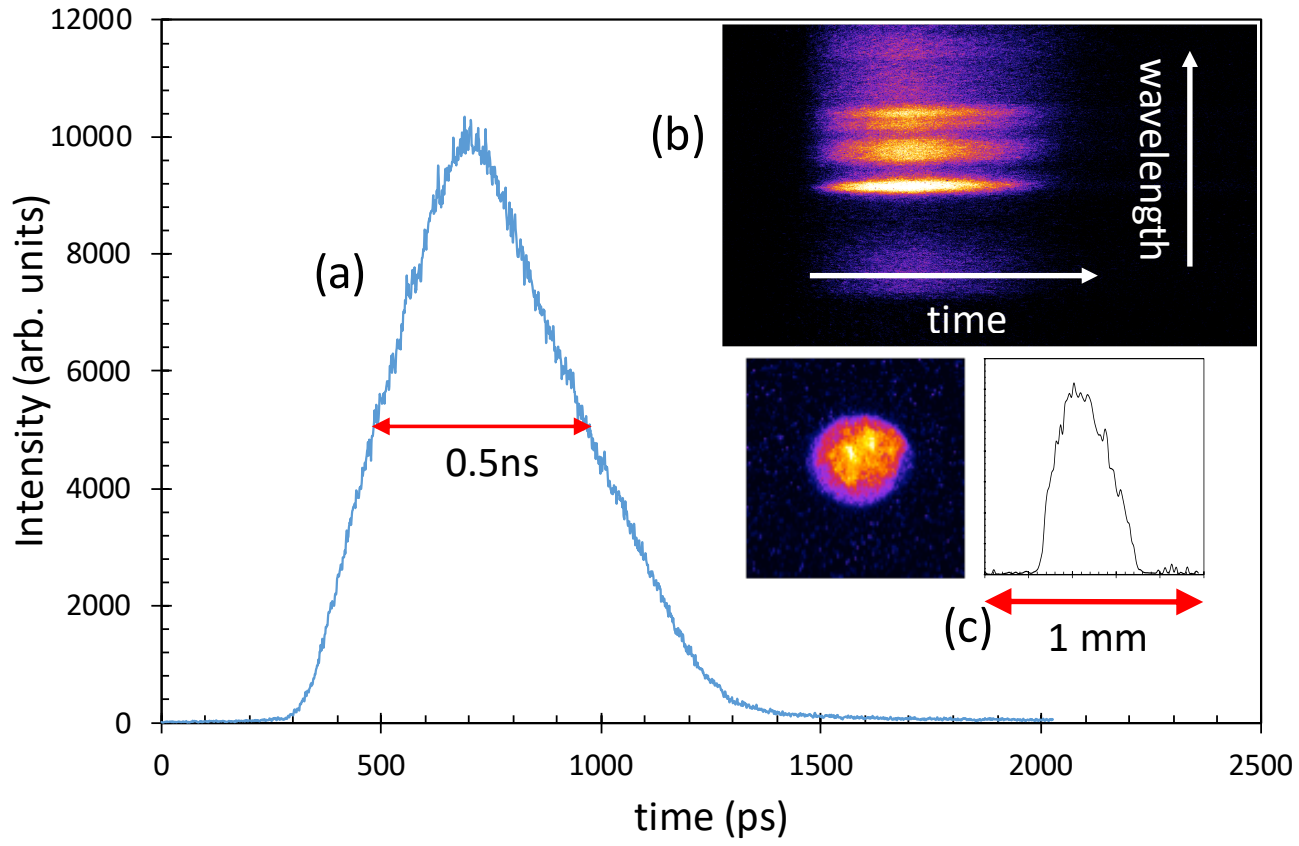

FIG. 3. (a) Lineout of the streaked X-ray emission for Sn L-band radiation measured with ~ 30 ps resolution. (b) Raw image of the X-ray streak data for the line group from ~3700-3850eV seen in Figure 2. The lineout in (a) shows the timescale (c) Raw X-ray pinhole image of the Sn plasma emission in the keV range and a vertical lineout. The FWHM of the X-ray emitting region is ~ 300 μm in this case.

Fig. 3 presents a streak image and lineout of the L-shell emission, which varied between shots from ~ 0.4–0.5 ns full-width-at-half-maximum (FWHM) duration. For each shot both time-resolved and time-integrated L-shell spectra and X-ray pinhole camera images were obtained.

## III. Results and discussion

Fig. 4(a) shows a raw image from the Mica crystal spectrometer [14] for a shot with 500 mbar gas pressure. Observation of the cold K-α doublet indicated a resolution of 1.2 eV FWHM. Fig. 4(a) also shows the emission from the X-ray drive plasma as it expands outwards away from the gas cell, while in Fig. 4(b) we present the image of a shot where the Sn layer was replaced with Al, which has much lower radiative emission in the 3 to 4 keV regime. There is virtually no K-α emission evident in this case, indicating that the production of the fluorescence was driven by radiative flux from the Sn plasma, and not by collisional excitation and ionization from plasma driven into the gas from the irradiated foil.

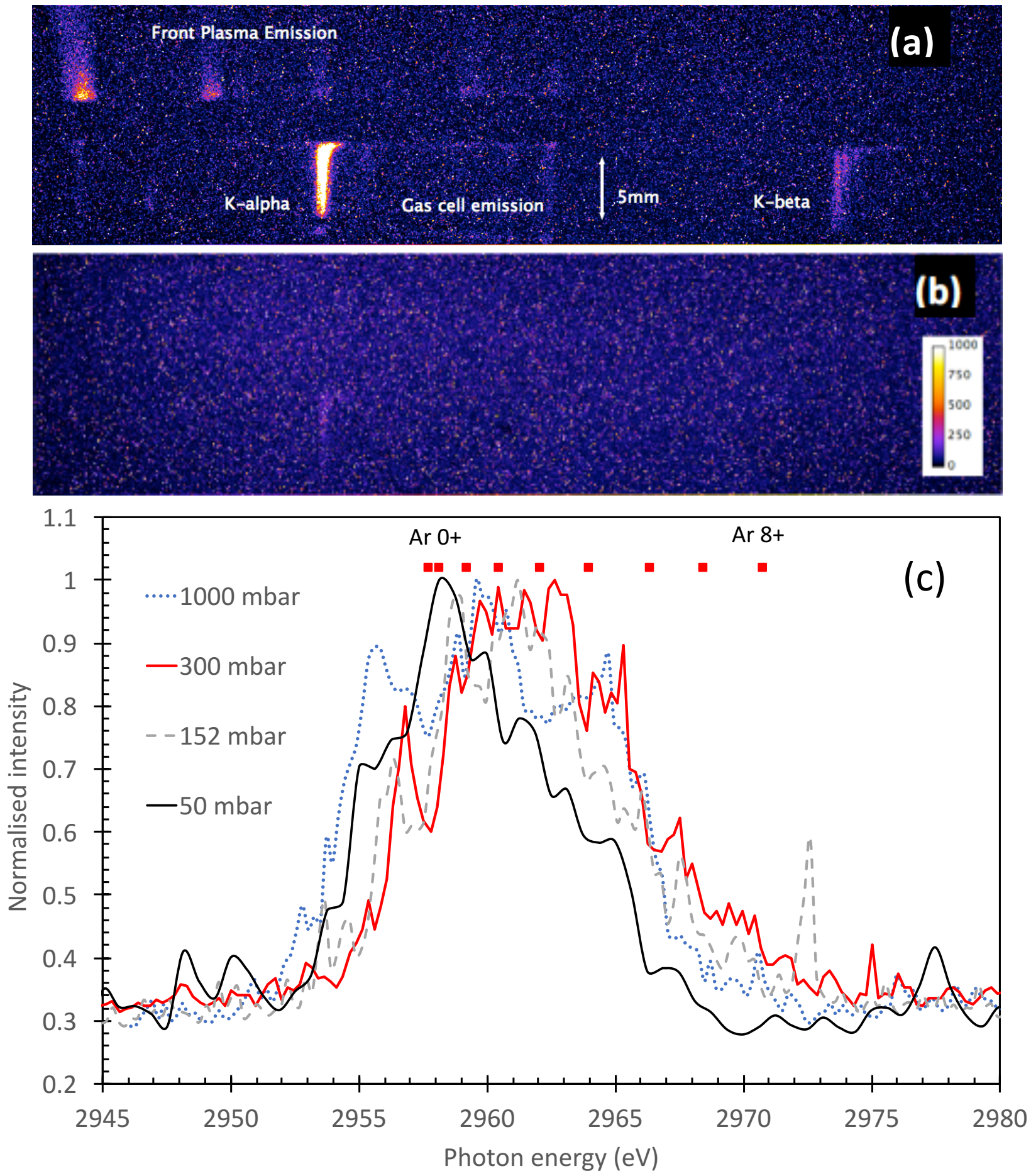

FIG. 4. (a) Raw CCD image for a 500 mbar Ar gas fill (678 J laser energy). A shadow region is evident between those where the Ar is viewed and where plasma from the expanding laser-driven material is visible as it moves away from the gas cell window. Hard X-rays from the Sn laser-plasma cause speckle across the chip. In (b) the shot at the same gas pressure (787 J laser energy) has the 218 nm Sn layer replaced with 186 nm Al, and there is a dramatic drop in the emission of Ar K-α and virtual disappearance of K-β emission. The calibration scale is the same for both images. (c) Raw spectra at varying gas pressure, taken ~0.5-0.7 mm from the X-ray drive source. For 1000 and 300 mbar there is a similar level of maximum ion stage observed, but as the pressure drops to 152 mbar and then 50 mbar the maximum ion stage also decreases. Red squares mark the expected positions of the Ar K-$\alpha_1$ lines from different ion stages, where the charge denoted refers to that of the initial ion prior to creation of the inner-shell vacancy by photoionization.

Raw spectra at varying gas pressure are presented in Fig. 4(c). The wavelength scale for this figure is set using the cold K-α emission far from the source and weak drive shots along with the position of the K-α to check the dispersion. There is only a small change in the highest ionization stages detected when the pressure drops from 1000 to 150 mbar and the spectra are similar. However, there is a more noticeable drop in average ionisation on a reduction to 50 mbar.

In Fig. 5(a) we show the spectral lineouts of despeckled images from two shots at 50 mbar, taken as close to the source as possible due to the geometry of the target (z = 0.5-0.7 mm). There is a shot-to-shot variation, which is worst for the lowest density case due to the signal level being proportional to the gas density. Nevertheless, the spectrum in both cases is centred on emission from 3 to 4 times ionized Ar. Assuming background Z* = 3 and that hydrodynamic timescales are longer than the duration of the X-ray drive, we can use the gas pressure to estimate an electron density of $3.6 \times 10^{18}$ $cm^{-3}$. We justify this in the light of estimates of the electron temperature below, which indicate the plasma sound speed should be ~ $10^4$ m $s^{-1}$. For a typical dimension of ~1 mm this implies a hydrodynamic timescale of ~ 100 ns. Fig. 5(b) shows a shot with gas pressure of 152 mbar, and we see that the spectrum is centred around a ~4 times ionized plasma. In both cases we compare with simulations, which are discussed below.

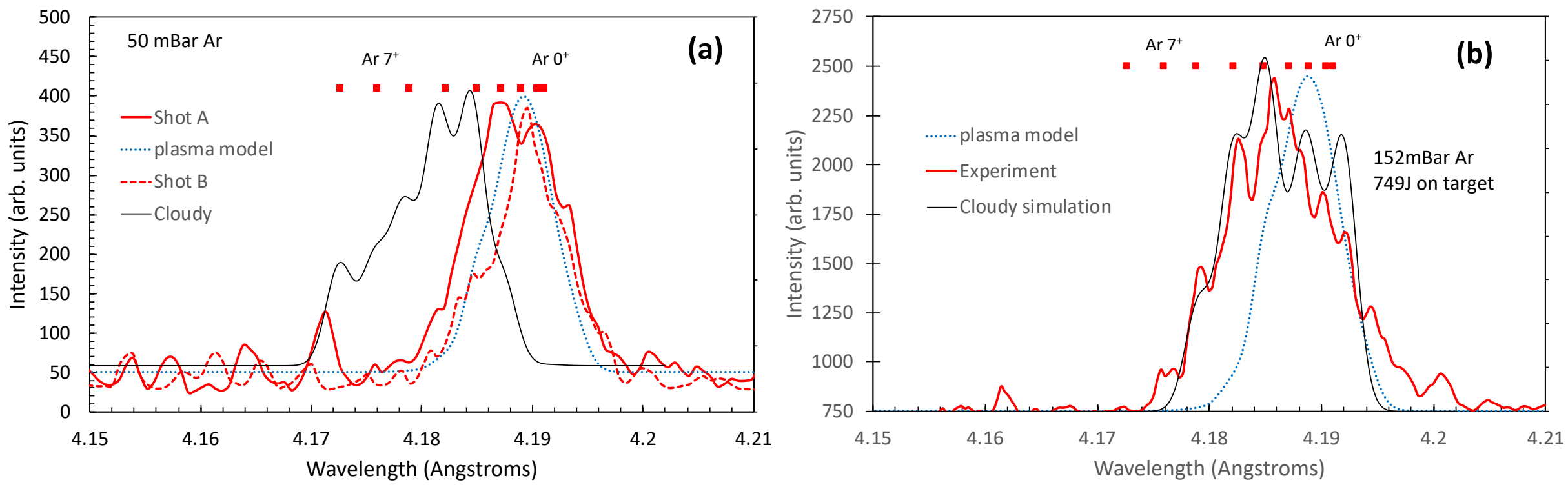

FIG. 5. (a) Lineout from two shots with a 50 mbar gas fill close to the source at z = 0.5-0.7 mm with incident flux ~$10^{19}$ ergcm$^{-2}$s$^{-1}$. The images are despeckled before the lineouts are taken (pixels replaced with median of a surrounding 3 x 3 grid, conserving overall signal). For this lowest pressure case shot-to-shot variation in the profile can be seen more strongly than at higher pressures due to lower signal levels. (b) Raw data lineout (0.5-0.7mm) from a shot with 152 mbar gas fill and a similar incident flux to (a). See the text for discussion of the simulations. In both figures we mark the positions of the expected K-$\alpha_1$ lines from different ion stages.

Using the efficiency of L-shell emission, we can estimate ξ. Our model accounts for the profile of the X-ray focal spot (fitted to a super-Gaussian) and assumes a $\cos(\theta)^{1/2}$ variation in X-ray emission [15]. This allowed us to model the emission as a function of the distance in the direction normal to the focal spot, and also in the direction parallel to the focal spot plane. The spectrometer spatially resolves in the direction normal to the Sn foil but spatially integrates in the parallel direction. Partial collimation resulting from use of an 0.5mm aperture on the 0.5 mm thick substrate means that, close to the source, there is a limited lateral spread in the source size. In the spatially resolved direction, our lineouts are averaged over an approximately 200 μm wide region. Taking into account the spatial averaging in both directions, we calculate that the effective weighted flux of L-shell radiation is ~$1.0 \times 10^{19}$ erg/cm$^2$/s at the peak of the pulse, with spatial variations of ±30%. For the 50 mbar case, at an electron density of $3.6 \times 10^{18}$ $cm^{-3}$ we have a weighted mean estimated value of ξ ~ 50 erg cm $s^{-1}$ (including the lower photon energy flux) at the peak of the flux. In Fig. 6 we show the geometry relevant to the calculation of the flux with our experimental arrangement.

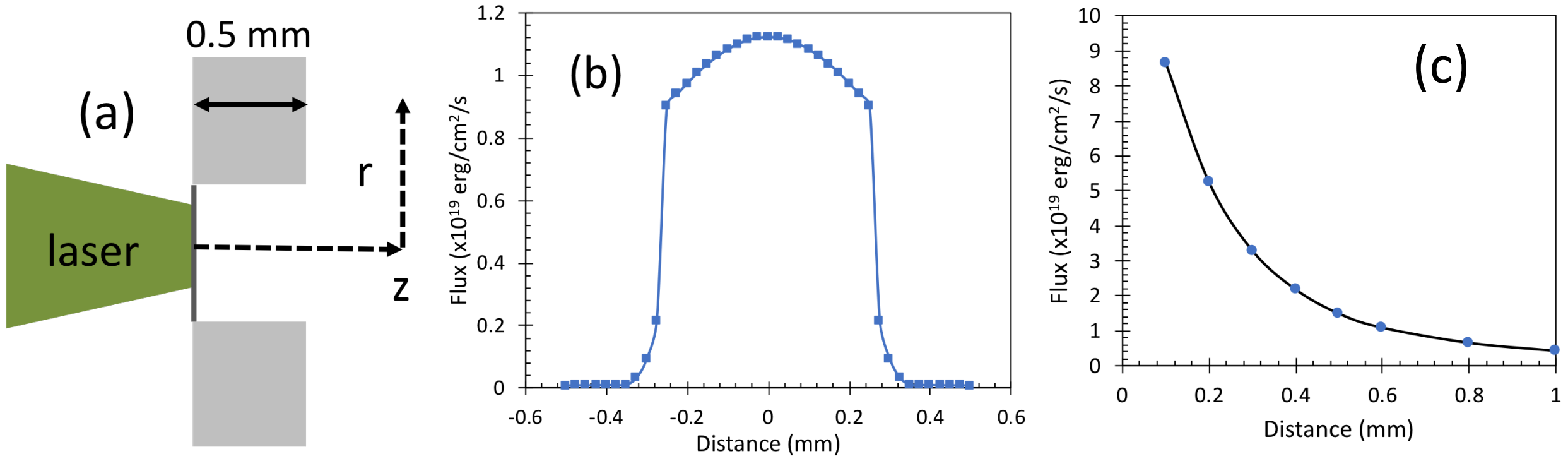

FIG. 6. (a) Geometry of the recessed target. The flux profile at the focal plane is taken from the pinhole data above. (b) profile of the peak flux calculated as a function of the distance, defined as r in part (a) of the figure, and at a distance z=0.6 mm. Away from the centre, the recess walls block flux and thus the flux is restricted in spatial extent and thus value. (c) Peak flux at r=0 calculated as a function of z. We can see there is a steep gradient in flux as discussed in the text.

To understand the expected behaviour of our plasma, we developed a simple time-dependent simulation model and assumed a uniform gas density for the duration of the X-ray drive. Collisional ionization data from Lotz [16] is fitted along with dielectronic recombination and radiative recombination rates by Shull and van Steenburg [17]. Shell-resolved photoionization cross-sections are from Verner and Yakovlev [18], and the fluorescence yields for K-shell vacancies from Kaastra and Mewe [19].

We use a weighted average photon energy for the L-shell of 3.8 keV. In addition, we model the contribution of lower energy photons by an assumed black-body of temperature 175±10 eV, passing through a cold CH foil as discussed above by dividing them into 7 photons groups up to 1.75 keV. This can be used to calculate the expected ratio of 1s and 2s photoionization rates from $Ar^{3+}$. Doing so leads to the same ratio as a black-body radiation temperature of $1.7^{+0.5}_{-0.3}$ keV. This is an important point of the technique as discussed by Hill and Rose [11]. Our model assumes that the fluorescence and Auger decay of ions with an inner-shell vacancy occurs instantaneously compared to the timescale of the experiment. The electron temperature is fixed by the energy balance between the absorbed energy from photoionization, the energy lost to radiative recombination, K-$\alpha$ emission and bremsstrahlung (see e.g. [20]), and the energy contained in the plasma due to ionization. As with the Auger and fluorescence timescales, we assume a rapid thermalisation time between electrons, which for an electron density of $>10^{18}$ cm$^{-3}$ and temperature of 15 eV (see below) would be sub-ps for Z*=3 [21]. The energies of the K-shell transitions were calculated with a Hartree-Fock model.

As seen in Fig. 5, for the 50 mbar shots, the emission predicted by our model is peaked at about the same ionization stage as the experimental data, while for the 152 mbar case the simulated ionization is clearly lower than that observed and this trend continues for higher pressures. For all pressures, the predicted spectrum is narrower than experiment despite the inclusion of spatial and temporal averaging in the simulation. In Fig. 7(a) we see the predicted rise in electron temperature as well as the emission history of K-$\alpha$ radiation for individual ion stages. An important parameter of a photoionized plasma is the degree to which photoionization dominates over collisional ionization. In figure 7(b) we show, for our weighted averaged conditions, the simulated temporal history for the ratio of total collisional to total photoionization rates for the $Ar^{3+}$ ion stage.

It can be seen that, photoionization is dominant early on, but as the plasma heats up, the collisional ionization rate rises to about 6 times that due to photoinization at the peak of the flux. Of course, for the inner-shell ionization, the photoionization is totally dominant over collisional ionization under these conditions.

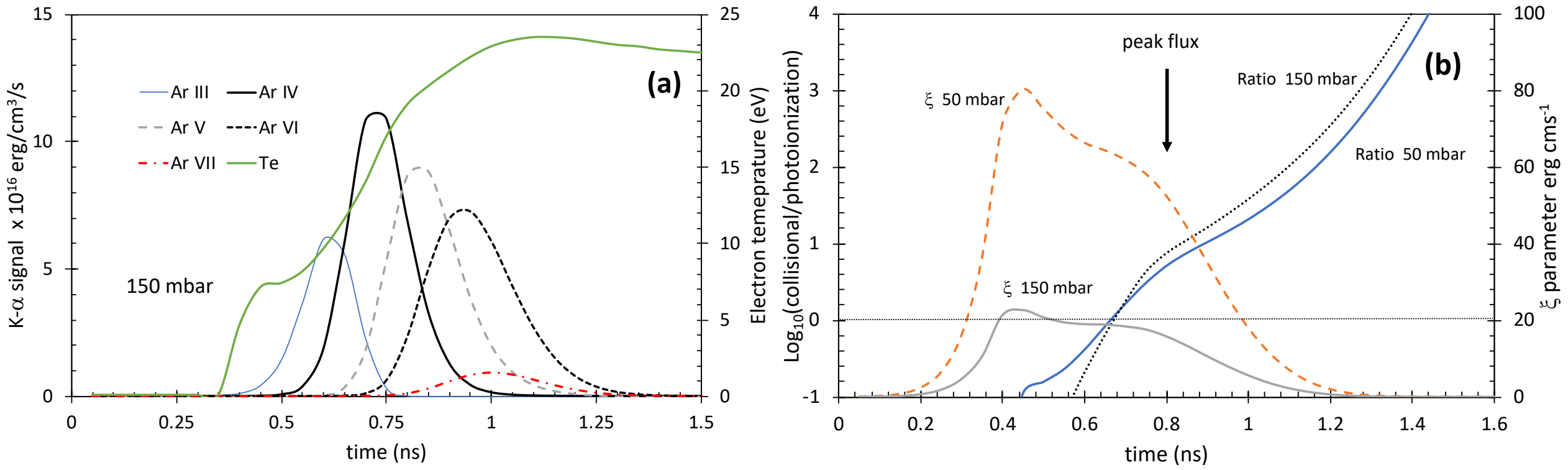

FIG. 7. (a) Calculated emission history using a simple plasma code as described in the text. The ion stages are those emitting the photons. The temperature rises to above 20 eV before radiative losses cool the plasma, the main emission predicted to occur when the temperature is 15-20 eV. (b) Temporal history of the ξ parameter for the lowest two pressures studied. The ratio of collisional ionization to photoionization for $Ar^{3+}$, shows that at early times prior to the peak flux we have a photoionization-dominated plasma.

Since one of our aims is ultimately to help benchmark codes relevant to astrophysical sources, we have compared our data with simulations from the Cloudy code [22,23]. This code is steady state and not expected to be strictly valid for our experiment but the comparison is interesting. For the lower pressure case it severely overestimates the expected ionisation, as might be expected. Agreement at higher density may partly be explained by faster equilibration, but a more detailed consideration of the timescales expected (see Fig. 6(a)) leads us to believe this is insufficient.

The fact that our time-dependent model predicts a lower ionization than the steady-state code is not surprising. However, it also predicts lower ionization than experiment and there are potential reasons we can consider. Firstly, there is the very low energy photon component (up to 500 eV) for which the outer-shell photoionization cross-section may be underestimated. However, the CH layer is thick and simulations suggest that during the drive laser pulse, there is little penetration of the heat front into the CH. A more likely scenario is energy transfer from the higher flux regions closer to the source (~$10^{20}$ erg/cm$^2$/sec for z=100 μm). Some of the energy absorbed will be re-emitted as M-band photons in the ~200-300 eV range as 2s and 2p vacancies are created by photoionization and Auger decay. Radiation transfer from high flux regions may increase the effective flux in adjacent plasma further from the focal spot.

For electrons at the predicted average temperature of ~20 eV for the 50 mbar case, the electron-electron equilibration time is of sub-ps duration. However, for the photoionized electrons, ejected with 0.5 to 1 keV energy, the electron-electron equilibration time is >50 ps for the 50 mbar case, with a collisional ionization time for reaching Z = 3 of ~20-30 ps. For these cases the mean-free-path of the electrons is of the order of several hundreds of microns and so we have to consider that energy can flow from higher to lower flux regions. These electrons would be below the threshold for K-α production but can ionize the outer shell electrons. Modelling of both the non-local electron transport and radiative transfer is a complex problem requiring detailed code development. However, we have attempted to estimate the effect of an incoming net flow of faster electrons from a higher flux region, closer to the Sn foil source by adding in a small percentage of 'hot' electrons. We have done this and the results in Fig. 8 show a stronger effect for higher pressure, which is consistent with our experimental observation that higher pressures do start to show higher average ionization.

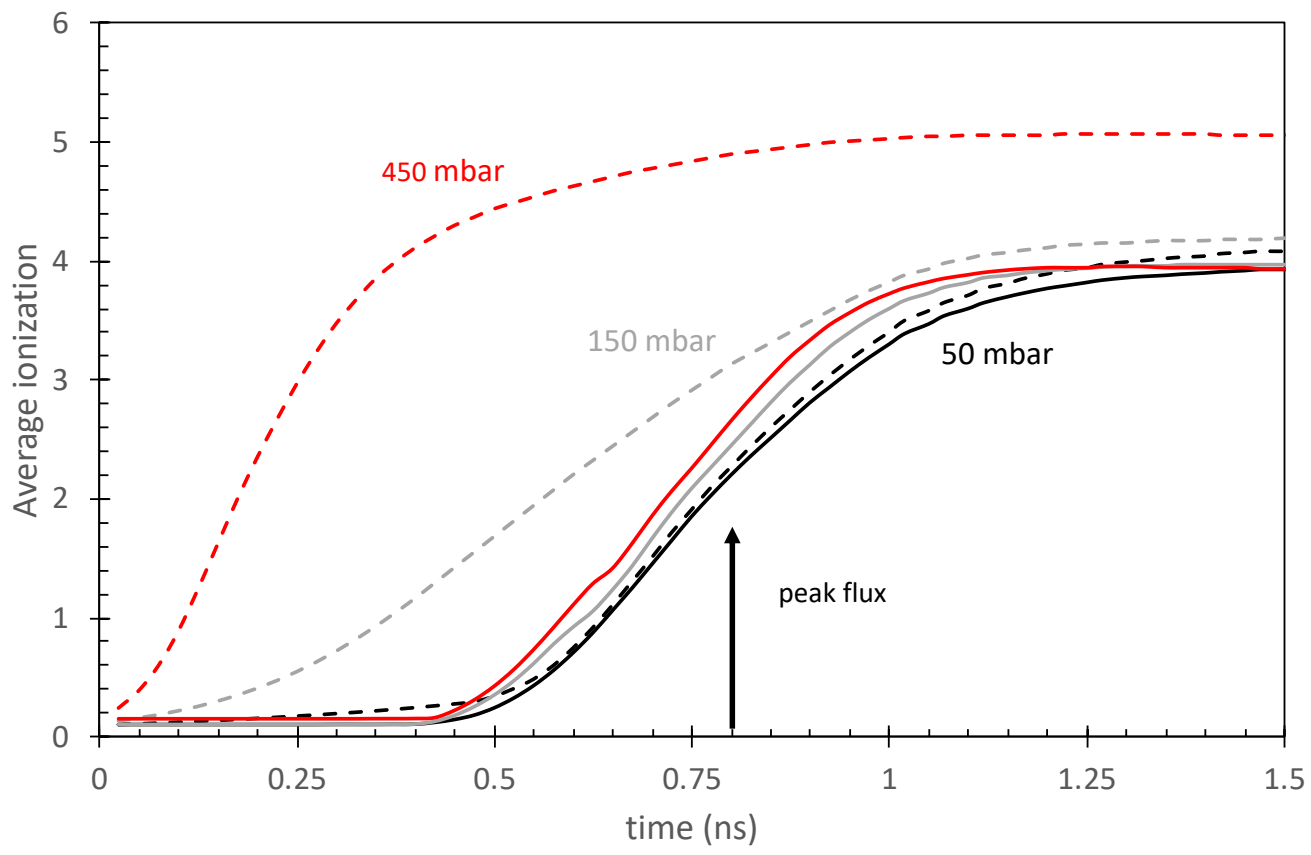

FIG. 8. In this plot we show the time history of average ionization for three gas pressures. The incident flux of L-shell radiation is $10^{19}$ ergs/cm$^2$/s and an additional 2% population of fast electrons is added assuming an effective temperature of 250eV. This adds about 25% to the energy deposited. More detailed modelling would be needed for a full understanding but this simple calculation shows that the effect is more pronounced for higher gas pressures. This is consistent with the higher ionisation observed for higher gas pressure in the data.

In comparing our experiment to previous work, we note that our drive duration is somewhat shorter than some cases [8,10] but longer than others [9], and the K-α fluorescence is not emitted from a steady-state plasma. The drive flux of Foord *et al* [8] is higher, and our value of $\xi$ is due to working at a lower electron density. Although a lower density is, in principle, more relevant to astrophysical plasmas, it means that the equilibration time between electrons and ions is longer than the drive duration in our case. We also note that for the spatially-averaged drive conditions, the total photoionization rate from the $Ar^{3+}$ ion stage is higher than the collisional ionization rate (50 mbar case) up until 150 ps before the peak flux, after which the collisional ionization rate rapidly comes to dominate. Thus, although the plasma is completely created by photoionization, much of the time-integrated emission comes from plasma where collisional processes are dominant. This is a common situation for any laboratory-based experiments, but will also be a potential feature of plasmas in fusion environments and thus our experiments can still provide a relevant test for codes under a range of conditions. Further simulations using our simple model have indicated that for a steady flux at $10^{19}$ erg cm$^{-2}$ s$^{-1}$, the rate of absorption of energy by photoionization would be balanced by losses due to bremsstrahlung and radiative recombination after ~ 20–30 ns, when the temperature would be much higher (> 150 eV) than in our non-steady-state experiment. On this timescale we may also reach equilibration in the electron and ion temperatures which for our conditions should be of the order of several ns. This is encouraging, as there are higher energy laser systems than VULCAN that would be capable of sustaining such an X-ray flux for durations of this order.

**Summary**

In summary, we have implemented a proposed experimental technique using a line radiation group to photoionize a plasma in order to mimic a ~keV temperature radiation field . We have achieved photoionization parameters of $\xi$ ~ 50 ergs cm s$^{-1}$. The small spatial scale of the experiment and low starting density has helped in achieving this result but also means that more detailed modelling (including electron and photon transport) is needed due to the steep gradient in flux. Extension to larger energy laser systems may allow experiments with double-sided irradiation and larger spatial scales, thus mitigating the gradient issues and helping to create plasmas that can be used to benchmark relevant codes.

## ACKNOWLEDGEMENTS

This work was supported by the UK Science and Technologies Facilities Council, the National Natural Science Foundation of China under grant No. 11573040 and Science Challenge Project No. TZ2016005. The authors gratefully acknowledge the expert support from the VULCAN laser facility. Supplementary data and raw images used are available via www.qub.ac.uk/pure.